\documentstyle[aps,epsf]{revtex}
\begin{document}
%\preprint{NSF-ITP-97-134, IASSNS-HEP 97/119}
%\widetext
\draft
\title{A Chern-Simons Effective Field Theory for the Pfaffian
Quantum Hall State}
\author{E.~ Fradkin$^{1}$, Chetan~ Nayak$^{2}$, A.~ Tsvelik$^{3}$,
Frank~Wilczek$^{4}$}
\address{
$^{1}$ Dept. of Physics, University of Illinois at Urbana-Champaign,
Urbana, IL, 61801\\
$^{2}$ Institute for Theoretical Physics,
University of California,
Santa Barbara, CA, 93106-4030\\
$^{3}$ Department of Theoretical Physics,
University of Oxford, 1 Keble Road,
OX1 3NP, U.K.\\
$^{4}$ School of Natural Sciences,
Institute for Advanced Study, Olden Lane,
Princeton, N.J. 08540}

\maketitle

\begin{abstract}
We present a low-energy effective field theory
describing the universality class of the
Pfaffian quantum Hall state.
To arrive at this theory, we observe that
the edge theory of the Pfaffian state of
bosons at $\nu=1$ is an $SU(2)_2$ Kac-Moody
algebra. It follows that the  
corresponding bulk effective
field theory is an $SU(2)$ Chern-Simons theory
with coupling constant $k=2$.
The effective field theories for other
Pfaffian states, such as the fermionic one
at $\nu=1/2$ are obtained by a flux-attachment
procedure.
We discuss the non-Abelian statistics of
quasiparticles in the context of this
effective field theory.
\end{abstract}

%\narrowtext

\section {Introduction}

Recently there has been considerable interest in a new class of
quantum Hall states, combinining aspects of
BCS pairing with Laughlin-type
ordering \cite{mooreread,greiter,wilczek,mr,rr,haldane}.
The states appear to be incompressible, and
to exhibit non-Abelian statistics.  Their properties have mainly been
inferred by extrapolation in quantum statistics (of electrons) 
from ordinary superconductors \cite{greiter},
from numerical studies \cite{greiter,haldane},
and from the existence of attractive trial
wave functions \cite{mooreread,greiter,wilczek,mr,rr}. 
To assure ourselves of the robustness of these
properties, and for calculational purposes, it is important to have a
low-energy, long-wavelength effective field theory.  In this paper, we
shall supply such a theory.

 Analyses of the  ground states of the lowest Landau level Hamiltonian 
with three-body interactions  
of the form
\begin{equation}
H = {\sum_{i>j>k}} \delta'({z_i}-{z_j})\delta'({z_i}-{z_k})
\end{equation}
have unearthed a number of fascinating
properties of these states \cite{mooreread,greiter,wilczek,mr,rr}.
In particular, at Landau level
filling fraction $\nu=1/2$, the ground state is
the so-called Pfaffian state,
\begin{equation}
{\Psi_{\rm Pf}}\,\, =\,\,
{\rm Pf}\left(\frac{1}{{z_i} - {z_j}}\right)
\,\,\,{\prod_{i>j}}{\left({z_i} - {z_j}\right)^2}\,\,\,
{e^{-\frac{1}{4{\ell}_0^2} \sum |z_i|^2 }}
\label{pfaffian}
\end{equation}
(where ${\rm Pf}\left(\frac{1}{{z_i} - {z_j}}\right) =
{\cal A}\left(\frac{1}{{z_1} - {z_2}}\,
\frac{1}{{z_3} - {z_4}}\,\ldots\right)$ is the antisymmetrized product
over pairs of electrons).
There is a gap to all excited states and a discontinuity
in the chemical potential, so (\ref{pfaffian}) describes
an incompressible state with quantized Hall conductance.
This state (or a state in the same universality class,
in a sense to be discussed below) might describe the
plateaus observed at $\nu=5/2$
in a single-layer system and $\nu=1/2$ in double-layer systems.

It will be useful to consider a slight generalization
of (\ref{pfaffian}):
\begin{equation}
{\Psi_{\rm Pf}}\,\, =\,\,
{\rm Pf}\left(\frac{1}{{z_i} - {z_j}}\right)
\,\,\,{\prod_{i>j}}{\left({z_i} - {z_j}\right)^q}\,\,\,
{e^{-\frac{1}{4{\ell}_0^2} \sum |z_i|^2 }}
\label{qpfaffian}
\end{equation}
For $q$ even, (\ref{qpfaffian}) is a state of fermions at $\nu=1/q$;
for $q$ odd,  (\ref{qpfaffian}) is a state of bosons. The bosonic
state with $q=1$ is, in fact, an approximate ground state
at $\nu=1$ of bosons interacting through the Coulomb
interaction \cite{haldane}.

As charge is removed from the Pfaffian state (\ref{pfaffian}),
quasiholes are created. The quasiholes carry half of a flux
quantum, as may be seen from the following state,
which has quasiholes at $\eta_1$ and $\eta_2$:
\begin{equation}
{\Psi_{\rm Pf}}\,\, =\,\,
{\rm Pf}\left(\frac{\left({z_i}-{\eta_1}\right)\left({z_j}-{\eta_2}\right)
+ i\leftrightarrow j}
{{z_i} - {z_j}}\right)
\,\,\,{\prod_{i>j}}{\left({z_i} - {z_j}\right)^2}\,\,\,
{e^{-\frac{1}{4{\ell}_0^2} \sum |z_i|^2 }}
\label{twoqhpfaffian}
\end{equation}
The most remarkable features of the Pfaffian state
are exhibited when there are at least four quasiholes.
A state with $4$ quasiholes at points $\eta_{\alpha}$
is obtained by modifying the Pfaffian as follows:
\begin{equation}
{\rm Pf~ }\Biggl( {1\over z_j - z_k }\Biggr) \rightarrow
{\rm  Pf~  }\Biggl( { (z_j - \eta_1) (z_j - \eta_2)(z_k
-\eta_3) (z_k - \eta_4) + (j \leftrightarrow k )
\over z_j - z_k}\Biggr)\equiv{{\rm Pf}_{(12)(34)}}~.
\label{fourqh}
\end{equation}
However, this is not the only four-quasihole state
with quasiholes at ${\eta_1},{\eta_2},{\eta_3},{\eta_4}$.
${{\rm Pf}_{(13)(24)}}$ and ${{\rm Pf}_{(14)(23)}}$
seem to be equally good. In fact, only two of these
three are linearly independent, as a consequence of the
identity \cite{wilczek}:
\begin{equation}
{{\rm Pf}_{(12)(34)}}-{{\rm Pf}_{(14)(23)}}=
{ {{\eta_{14}}{\eta_{23}}}\over{{\eta_{13}}{\eta_{24}}} }
\,\,\biggl({{\rm Pf}_{(12)(34)}}-{{\rm Pf}_{(13)(24)}}
\biggr)~.
\label{fourqhiden}
\end{equation}
It can similarly be shown that there are $2^{n-1}$
linearly independent states of $2n$ quasiholes
at fixed positions ${\eta_1},{\eta_2},\ldots,{\eta_{2n}}$ \cite{wilczek}.
This exponential degeneracy has been interpreted in
terms of the possible occupation numbers of $n$
fermionic zero modes associated with the
quasiholes \cite{mr,rr,haldane}. As follows from the numerical
value of the degeneracy, the above
mentioned zero modes are Majorana fermions.
We will see below that these fermions appear as
edge states in a system of finite size.

The degeneracy of the multi-quasihole states allows for the possibility
of non-Abelian statistics. This possibility is, indeed, realized,
and the braiding matrices are associated
with the group $SO(2n)$ \cite{wilczek}. This remarkable relationship
is possible because $SO(2n)$ has a $2^{n-1}$ dimensional
spinor representation. Here we again encounter Majorana fermions
since the spinor representation is constructed from 2n
real $\gamma$-matrices satisfying the fermionic anticommutation
relations
\begin{equation}
\{\gamma_i, \gamma_j\} = \delta_{i,j}
\end{equation}
The operation of braiding the $i^{th}$
and $j^{th}$ quasiholes corresponds (up to a phase) to the $SO(2n)$ rotation
by $\pi$ in the $i-j$ plane.
In other words, the action
of this braiding operation on the space of $2n$ quasihole states
is given by the matrix which
represents the corresponding $SO(2n)$ rotation in
its spinor representation. For instance, an exchange
of $\eta_1$ and $\eta_3$ leads to the following
rotation between two four-quasihole states
(in a particular basis of the four-quasihole states
whose definition is unimportant here)
\begin{equation}
\label{wavebraidmat}
\frac{e^{i\pi\left(\frac{1}{8}+\frac{1}{4q}\right)}}{\sqrt{2}}\,
\left(\matrix{1& 1\cr -1  &  1}\right)
\end{equation}
This is precisely the matrix in the spinor
representation of $SO(4)$ which represents a
rotation by $\pi$ in the $1-3$ plane.

Another important topological quantum number characterizing a quantum
Hall state is the ground state degeneracy on a torus.
At $\nu=1/q$, there is always a trivial $q$-fold
center-of-mass degeneracy coming from the $\theta$-function
generalization of the Jastrow factor. The Pfaffian state
has ground state degeneracy $3q$, with the additional
factor of $3$ coming from the toroidal version
of the Pfaffian:
\begin{equation}
{\rm Pf}\left(\frac{1}{{z_i} - {z_j}}\right) \,\rightarrow\,
{\rm Pf}\left(\frac{{\theta_a}({z_i} - {z_j})}
{{\theta_1}({z_i} - {z_j})}\right)
\end{equation}
where $a=2,3,4$.

The edge excitations of the Pfaffian state
take the form \cite{mr,rr}:
\begin{equation}
{\cal A}\left({{z_1}^{p_1}}\ldots{{z_k}^{p_k}}
\frac{1}{{z_{k+1}} - {z_{k+2}}}\,
\frac{1}{{z_{k+3}} - {z_{k+4}}}\,\ldots\right)\,\,
S({z_1},\ldots,{z_N})
\,\,\,{\prod_{i>j}}{\left({z_i} - {z_j}\right)^2}\,\,\,
{e^{-\frac{1}{4{\ell}_0^2} \sum |z_i|^2 }}
\label{pfaffianedge}
\end{equation}
The arbitrary distinct positive integers ${p_1}, \ldots, {p_k}$
correspond to the creation of neutral fermionic excitations;
this is the $c=1/2$ sector of the edge theory.
$S({z_1},\ldots,{z_N})$ is a symmetric polynomial; this modification of the
ground state corresponds to the creation of chiral bosonic excitations in
the $c=1$ charged sector of the edge theory. At $\nu=1/2$,
the compactification radius of the boson is $R=1/\sqrt{2}$.
It can be shown that these excitations span the Hilbert
space of the edge theory; hence the edge theory
has total central charge $c=\frac{1}{2} + 1$.

The specific form of the trial wavefunctions considered
above played a crucial role in the determination of
the properties of the Pfaffian state,
such as the non-Abelian braiding statistics. Of course, we would
like to believe that the Pfaffian state is a representative of
an entire universality class of states which have the same
`topological properties', such as braiding statistics and
ground state degeneracy on the torus. To investigate the
existence and stability of this universality class, we need
a low-energy, long-wavelength effective field theory
for the Pfaffian state.

An effective field theory for the Laughlin states at $\nu=1/(2k+1)$
was obtained by a Landau-Ginzburg construction
\begin{eqnarray}
{{\cal L}_{\rm eff}} = {\psi^*}\left(i{\partial_0}-
\left({a_0}+{A_0}\right)\right)\psi &+&
\frac{\hbar^2}{2m^*}\,{\psi^*}{\left(i\nabla -
\left({\bf a}+{\bf A}\right)\right)^2} \psi+
u{|\psi|^4}\cr
&+&\, \frac{1}{2k+1}\,\frac{1}{4\pi}
{\epsilon^{\mu\nu\rho}}{a_\mu}{\partial_\nu}{a_\rho}
\label{langinz}
\end{eqnarray}
The fundamental objects in the Landau-Ginzburg theory are auxiliary
bosons, $\psi$, which are electrons with fictitious flux attached
via ${\bf a}$. The
fractionally charged, anyonic quasiparticles are vortices.
In a dual theory, which results from integrating out
the auxiliary bosons, the quasiparticles are fundamental:
\begin{equation}
{{\cal L}_{\rm dual}} = \frac{2k+1}{4\pi}
{\epsilon^{\mu\nu\rho}}{\alpha_\mu}{\partial_\nu}{\alpha_\rho}+
{A_\mu}{\epsilon^{\mu\nu\rho}}{\partial_\nu}{\alpha_\rho}+
{\alpha_\mu}{j_{\rm vortex}^\mu}
\label{duallag}
\end{equation}
This dual theory has a simple relationship with the edge theory,
and seems more amenable to a non-Abelian generalization.

We do not know what the correct Landau-Ginzburg theory is
for the Pfaffian state (see, however, our comments at the end).
However, we do know that the edge theory is the conformal
field theory of a Majorana fermion and a free boson,
with $c=\frac{1}{2} + 1$. We will use this edge theory to
deduce the effective field theory of the bulk (which is
dual to the Landau-Ginzburg theory). Our first step will
be to show how this is done in the simplest
case (for reasons which will become clear) of a
Pfaffian state of bosons at $\nu=1$. We will then check that
this effective field theory reproduces the bulk properties
exhibited by the wavefunctions. Finally, we will
comment on the stability of the state.

\section{Effective Field Theory of the
Bosonic Pfaffian State at $\nu=1$}

The edge theory of the bosonic Pfaffian state at $\nu=1$
has $c=\frac{1}{2} + 1$, but the compactification radius
of the $c=1$ sector is $R=1$.
A marvelous feature of the free bosonic theory at $R=1$
is that it can be fermionized. The boson, $\phi$, can be replaced
by a Dirac fermion, $\psi$, or, equivalently,
two Majorana fermions, ${\chi_1}, {\chi_2}$:
\begin{equation}
{e^{i\phi}} = \psi = {\chi_1}+ i {\chi_2}
\end{equation}
Hence, the edge theory is a theory of a triplet of Majorana
fermions, with central charge $c=3/2$.

The triplet of Majorana fermions tranform under the
spin-$1$ representation of $SU(2)$. Let us call our
third Majorana fermion ${\chi_3}$.
The currents (the $T^a$'s are $SU(2)$ generators
in the spin-$1$ representation),
\begin{equation}
{J^a} = {\chi_i}{\left(T^a\right)_{ij}}{\chi_j}
\end{equation}
form an $SU(2)$ Kac-Moody algebra at level $k=2$.
\begin{equation}
{J^a}(z){J^b}(0) \sim \frac{2{\delta^{ab}}}{z^2} +
\frac{{f^{abc}}{J^c}(0)}{z} + \ldots
\end{equation}
The $U(1)$ subgroup of $SU(2)$ which is generated by
${J^3}$ is the $U(1)$ of electric charge since
${J^3}=  i{\chi_1}{\chi_2} = i\partial\phi$.

This Kac-Moody algebra has a bosonic incarnation
as the  {\it chiral sector} of the $SU(2)$ WZW model at $k=2$:
\begin{equation}
S = \frac{2}{16\pi}\,
\int\,{d^2}x\,tr\left({g^{-1}}{\partial_\mu}g\,{g^{-1}}{\partial^\mu}g\right)
+ \frac{2}{24\pi}\, \int\,{d^3}x\, {\epsilon^{\mu\nu\lambda}}
\,tr\left({g^{-1}}{\partial_\mu}g\,{g^{-1}}{\partial_\nu}g
\,{g^{-1}}{\partial_\lambda}g\right)
\end{equation}
where $g$ is an $SU(2)$-valued matrix  field.
The $SU(2)$ WZW model has $c=\frac{3k}{k+2} = 3/2$; which is 
the central charge of a  triplet of Majorana
fermions which indicates the  equivalency between  these two models. 
Hence, we may take the chiral sector of the $SU(2)_2$ WZW model
as the edge theory of the bosonic Pfaffian state at $\nu=1$.

Now, we are most of the way home because the bulk theory corresponding
to the chiral $SU(2)_2$ WZW model is simply the  $SU(2)$ Chern-Simons
theory with Chern-Simons coefficient $k=2$ \cite{witten},
\begin{equation}
S = \frac{2}{4\pi} \,\int\,{d^3}x\,{\epsilon^{\mu\nu\lambda}}\,
\left({a_\mu^a}{\partial_\nu}{a_\lambda^a} + \frac{2}{3}
{f_{abc}}{a_\mu^a}{a_\nu^b}{a_\lambda^c}\right)
\label{bosoniceft}
\end{equation}
where ${a_\mu^a}$ is an $SU(2)$ gauge field.
An alternative approach begins with the coset
construction of the $c=1/2$ theory
\cite{mooreseiberg,kogan}, but it is less transparent,
so we do not pursue it further.

\section{Topological Properties in the Effective Field Theory}

We can now check that the effective field theory
(\ref{bosoniceft}) predicts the same topological
properties as the  analysis of trial wavefunctions.
Let us first consider the ground state degeneracy on a 
torus, which we know to be three for bosons at $\nu=1$.
The Hilbert space of the Chern-Simons theory
(\ref{bosoniceft}) (since the Chern-Simons Hamiltonian vanishes,
all states are ground states) on a torus
can be obtained in the following way \cite{witten,verlinde}.
Consider the path integral
of (\ref{bosoniceft}) over the three-dimensional
region $M$ enclosed by a torus, $\partial M = {T^2}$,
\begin{equation}
\Psi[a] = {\int_{\alpha_{|T^2}=a}}\,\, D\alpha \,\,
{e^{2 \,{\int_M}\,{d^3}x\,{\epsilon^{\mu\nu\lambda}}\,
\left({\alpha_\mu^a}{\partial_\nu}{\alpha_\lambda^a} + \frac{2}{3}
{f_{abc}}{\alpha_\mu^a}{\alpha_\nu^b}{\alpha_\lambda^c}\right)}}
\end{equation}
subject to the condition that the gauge field $\alpha_\mu^a$
is equal to a prescribed gauge field $a_\mu^a$
at the toroidal boundary of $M$.
This is a state $\Psi[a]$ in the Hilbert space of (\ref{bosoniceft}).
Other states can be obtained by inserting a Wilson loop
in the path integral. Only a Wilson loop that is
non-contractable in $M$ -- there is only one such topologically
distinct loop -- will give a non-trivial contribution.
Furthermore, at level $k=2$, only Wilson loops in the
$SU(2)$ representations $j=0,1/2,1$ will give
non-vanishing path integrals \cite{witten}.
Hence, we have the three ground states on the torus:
\begin{equation}
{\Psi_j}[a] = {\int_{\alpha_{|T^2}=a}}\,\, D\alpha \,\,\,
{{\rm Tr}_j}\left\{{\cal P}{e^ {i\oint a}}\right\}\,\,\,
{e^{\frac{2}{4\pi} \,{\int_M}\,{d^3}x\,{\epsilon^{\mu\nu\lambda}}\,
\left({\alpha_\mu^a}{\partial_\nu}{\alpha_\lambda^a} + \frac{2}{3}
{f_{abc}}{\alpha_\mu^a}{\alpha_\nu^b}{\alpha_\lambda^c}\right)}}
\end{equation}
where ${{\rm Tr}_j}$ is the trace in the spin $j$ representation
of $SU(2)$ and ${\cal P}$ denotes path-ordering.

To obtain the degeneracy of the $2n$ quasihole states,
we need to first observe that the half-flux-quantum quasiholes carry
the spin-$1/2$ representation of $SU(2)$. Let's consider the
four quasihole case; the extension to $2n$ quasiholes is straightforward.
We would like the Hilbert space of (\ref{bosoniceft}) with
four external charges carrying the spin-$1/2$ representation of $SU(2)$.
At large $k$ (weak coupling), it is clear that these charges
do not interact except through the constraint that
the state be a total $SU(2)$ singlet. There are two different
spin-singlet states which can be made with four spin-$1/2$'s:
$(|+\rangle |-\rangle \,-\, |-\rangle |+\rangle)
(|+\rangle |-\rangle \,-\, |-\rangle |+\rangle)$
and $|+\rangle |+\rangle |-\rangle |-\rangle \,+\,
|-\rangle |-\rangle |+\rangle |+\rangle \,-\,
(|+\rangle |-\rangle \,+\, |-\rangle |+\rangle)
(|+\rangle |-\rangle \,+\, |-\rangle |+\rangle)$.
Similary, there are $2^{n-1}$ such states that
can be made with $2n$ spins subject to the constraint that
no subset of the spins can form a conglomerate
with spin greater than $1$. A more sophisticated analysis
shows that all $k>1$ are effectively
in the large-$k$ limit \cite{witten}.

These braiding eigenvalues can be derived in an appealing way from the
Chern-Simons theory.
This approach begins with the observation \cite{witten}
that the functional integral of our effective field
theory with Wilson loop insertions in the spin-$1/2$
representation is equal
to the Jones polynomial of the loops
evaluated at $q= {e^{\pi i/4}}$.
\begin{equation}
\int\,\, D\alpha \,\,\,
{{\rm Tr}_{1/2}}\left\{{\cal P}{e^ {i{\oint_\gamma} a}}\right\}\,\,\,
{e^{\frac{2}{4\pi} \,\int\,{d^3}x\,{\epsilon^{\mu\nu\lambda}}\,
\left({\alpha_\mu^a}{\partial_\nu}{\alpha_\lambda^a} + \frac{2}{3}
{f_{abc}}{\alpha_\mu^a}{\alpha_\nu^b}{\alpha_\lambda^c}\right)}}
\,\,\,=\,\,\,{V_\gamma}({e^{\pi i/4}})
\end{equation}
\begin{figure}
%\centerline{\psfig{figure=skein3.eps,height=1.5in}}
%\vskip 0.5cm
\vspace{.2cm}
\noindent
\hspace{1.5 in}
\epsfxsize=3.5in
\epsfbox{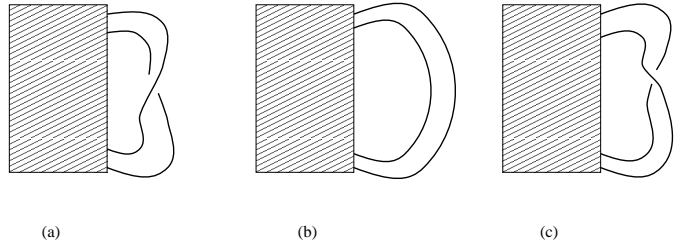}
\vspace{.5cm}
\caption{The loops (a) $\gamma$ (b) $\gamma'$ and (c) $\gamma''$
which enter the skein relation. The three loops differ only
by the braiding shown. The common shaded section
is arbitrary.}
\label{skein3}
\end{figure}
For our purposes,
the Wilson loops represent the world-lines of quasiparticles
and the Jones polynomial will tell us how quantum-mechanical
amplitudes depend on the braiding of the world-lines.

Let us recall a few facts about the Jones polynomial,
${V_\gamma}(q)$.
It is a Laurent series in one variable, $q$, which
is a topological invariant of a knot, $\gamma$. From
the point of view
of Chern-Simons theory, it is the next in
a hierarchy of invariants whose first member is the linking number,
which determines the acquired phase in Abelian
statistics. It is defined by the statement that
${V_\gamma}(q)=1$ if $\gamma$ is the unknot
and by the skein relation:
\begin{equation}
{q^{-1}}\,{V_\gamma}(q) - q\,{V_{\gamma''}}(q)
= \left({q^{1/2}} - {q^{-1/2}}\right){V_{\gamma'}}(q)
\label{skein}
\end{equation}
where $\gamma'$ and $\gamma''$ are obtained by
performing successive counter-clockwise
half-braids of any two world-lines in $\gamma$
as in figure 1.

Since (\ref{skein}) tells us how amplitudes
are modified by braiding operations, it
gives us a direct handle on the eigenvalues of the
braiding matrix. Consider an arbitrary state $|\psi\rangle$
in the two-dimensional space of states with four quasiholes.
According to the skein relation,
\begin{equation}
{q^{-1}}\,|\psi\rangle - q\,{B^2}|\psi\rangle
= \left({q^{1/2}} - {q^{-1/2}}\right) B|\psi\rangle
\label{skeinapplied}
\end{equation}
where $B$ is the braiding operator for two of the quasiparticles.
(\ref{skeinapplied}) implies a quadratic equation for the eigenvalues
of $B$ which yields the eigenvalues $e^{-3\pi i/8}$ and
$e^{\pi i/8}$, again in agreement
with (\ref{wavebraidmat}).

We can also obtain the quasiparticle statistics from the
braiding matrices of the spin-$1/2$ vertex operators
in the $SU(2)_2$ WZW model. According to \cite{mooreseiberg},
these follow from the fusion rules and anomalous dimensions in that
conformal field theory.

\section{Effective Field Theory of the
$\nu=1/2$ Pfaffian State}

We would now like to deform the theory (\ref{bosoniceft})
so that it describes the fermionic Pfaffian state at $\nu=1/2$.
(Of course, by the same procedure, we could also obtain
the effective field theory for fermionic Pfaffian
states at any even denominator or for bosonic Pfaffian
states at any odd denominator.) The idea is to use
the flux-attachment procedure \cite{aswz,zhk,fetter,lf} 
to change the bosons into
fermions and simultaneously change the filling fraction
from $\nu=1$ to $\nu=1/2$. To do this, we introduce a
$U(1)$ gauge field, $c^\mu$, which couples to the charge current,
${\epsilon^{\mu\alpha\beta}} {F^3_{\alpha\beta}}$. The field $c^\mu$
will attach a flux tube to each electron through the
Chern-Simons equation:
\begin{equation}
{\epsilon^{\mu\alpha\beta}}{\partial_\alpha} {c_\beta} = - {j^\mu}
= {\epsilon^{\mu\alpha\beta}} {F^3_{\alpha\beta}}
\end{equation}
This equation follows from the action
\begin{equation}
S = \frac{2}{4\pi} \,\int\,{d^3}x\,{\epsilon^{\mu\nu\lambda}}\,
\left({a_\mu^a}{\partial_\nu}{a_\lambda^a} + \frac{2}{3}
{f_{abc}}{a_\mu^a}{a_\nu^b}{a_\lambda^c}\right)
+ \frac{1}{2\pi} {c_\mu}{\epsilon^{\mu\alpha\beta}} {F^3_{\alpha\beta}}
- \frac{1}{4\pi} {c_\mu}{\epsilon^{\mu\alpha\beta}}{\partial_\alpha} {c_\beta}
\label{fermioniceft}
\end{equation}
which is our proposed effective field theory for
the fermionic Pfaffian state at $\nu=1/2$.

Equation (\ref{fermioniceft}) needs some explanation
because the term which couples ${a_\mu^a}$ and
${c_\mu}$ breaks the $SU(2)$ gauge invariance
down to the smaller $U(1)$ subgroup generated by
$T_3$ (as does the coupling to the electromagnetic field,
which has the same form). The action (\ref{bosoniceft})
and the equations of motion which
follow from it
are invariant under the transformation
\begin{equation}
{a_\mu^a}\,{T_a} \rightarrow g\,{a_\mu^a}\,\,{T_a}{g^{-1}}
- {\partial_\mu}g\,{g^{-1}}
\label{gaugetransf}
\end{equation}
where $g$ is an $SU(2)$-valued function
which must become  identity at the boundary.
As a result of (\ref{gaugetransf}),
the time-evolution of ${a_\mu^a}$ (in the bulk)
is not well-defined
because a given set of initial conditions
can lead to infinitely many possible solutions,
each related to the others by the  gauge transformation
(\ref{gaugetransf}). To put it differently, the
action does not specify a dynamics for the
longitudinal part of the gauge field, which
decouples from the transverse part.
On the other hand, gauge-invariant
quantities, such as ${F^a_{\alpha\beta}}{F^a_{\mu\nu}}$
have perfectly well-defined time-evolution
since they are independent of the longitudinal
part of ${a_\mu^a}$. For calculational
purposes, we can choose a particular gauge, thereby
specifying a dynamics for the longitudinal
part of ${a_\mu^a}$. Of course, gauge-invariant quantities
will not depend on this choice.

In (\ref{fermioniceft}), however, the
${c_\mu}{\epsilon^{\mu\alpha\beta}} {F^3_{\alpha\beta}}$
term breaks gauge-invariance by coupling the
transverse part of the gauge field to the longitudinal
part. As a result of its coupling to the longitudinal
part of ${a_\mu^a}$, the transverse part no
longer has a well-defined time-evolution either.
We can only make sense of the action
(\ref{fermioniceft}) if we understand the
$SU(2)$ part of the action to be gauge-fixed.
One possible choice is Coulomb gauge,
${a_0^a}=0$, or the weaker condition
${a_0^1}={a_0^2}=0$ which preserves the
$U(1)$ gauge symmetry generated by $T_3$.
In this gauge, there is a constraint,
\begin{equation}
{F_{12}^a}=0
\end{equation}
which generates time-independent gauge transformations.
We can impose another condition to eliminate this
residual gauge symmetry, but we do not have to.
Once we have imposed Coulomb gauge, a given set of
initial conditions leads to a unique solution.
The time-independent gauge transformations
connect different gauge-equivalent initial
conditions, but do not render the time evolution ill-defined.
In other words, we have a hugely redundant, but
completely well-defined dynamics.
Hence, we can take (\ref{fermioniceft}) as our
effective field theory, provided we understand the
$SU(2)$ gauge field to be gauge-fixed in the Coulomb
gauge or another suitable gauge.

Alternatively, and more profoundly, we can view the
${c_\mu}{\epsilon^{\mu\alpha\beta}} {F^3_{\alpha\beta}}$ term as a
gauge-fixed form of the gauge invariant operator ${\cal L}_{\rm spin}$
\begin{equation}
{\cal L}_{\rm spin} \equiv {\frac{1}{2\pi}}
\left(c_\mu+\partial_\mu \omega\right)
\epsilon^{\mu \alpha \beta}
F^a_{\alpha \beta} \phi^a
\label{operator}
\end{equation}
which is manifestly gauge invariant under both $U(1)$ and $SU(2)$ gauge
transformations. The scalar field $\omega$ is chosen to transform under an
arbitrary $U(1)$ gauge transformation $\alpha(x)$ as
$\omega(x) \to \omega(x)-\alpha (x)$, and the scalar field $\phi^a$
transforms like a vector of the adjoint representation of the $SU(2)$ gauge
group, and, as such, it transforms like $\phi^a(x) T^a \to g(x)\phi^a(x) T^a
g^{-1}(x)$,  where $g(x) \in SU(2)$. Then, after choosing the {\sl unitary}
gauge $\omega=0$ and $\phi^a=(0,0,1)$, we recover gauge non-invariant term of
(\ref{fermioniceft}).

Thus, we conclude that the effective field theory
must contain a scalar field in the adjoint representation of $SU(2)$. The
appearance of a scalar field in the adjoint representation of
$SU(2)$ in the effective low-energy theory indicates that the microscopic
quantum mechanical ground state must be characterized by a {\sl condensate} or
order parameter field which also transforms like a vector of the adjoint
representation of $SU(2)$. Therefore, since this vector also breaks the global
$SU(2)$ symmetry, the ground state must exhibit {\it spontaneous spin
polarization}. Furthermore, since the scalar field $\omega$ breaks the $U(1)$
gauge symmetry, we must conclude that this effective field theory describes a
state with off-diagonal long range order with a spin-triplet condensate. In
other words, the effective field theory describes a state with a $p-wave$
condensate. This conclusion is in perfect agreement with the arguments of
ref.~\cite{greiter}.

Before leaving this subject, we should add one more improvement.
Above, $\phi^a$ and $\omega$ have been  introduced as purely formal 
objects.  To  treat them  
as proper physical variables, we should impose  equations
of motion corresponding to their  variation.  If we did that using
(\ref{operator}) as it stands, we would obtain unsatisfactory
constraint equations.  We would also like to link these variables to
more familiar order parameter fields. So we introduce two charged
order parameter fields 
\begin{equation}
\Delta_0 = |\Delta_0|e^{i\omega}, ~~ \Delta^a = e^{i\omega}\phi^a
\end{equation} 
governed by the potential term 
\begin{equation}
{\cal L}_{\rm pot} ~=~ -
\lambda_1 ({\bf \Delta}{\bf \Delta}^+ -1 )^2 -
\lambda_2(\Delta_0\Delta_0^+ - 1)^2
\label{potential}
\end{equation}
with large parameters $\lambda_{1,2}$, thus fixing the amplitude of
these fields.  The Lagrangian density (\ref{operator}) then acquires a
more familiar form:
\begin{equation}
{\cal L}_{\rm spin} =  {\frac{1}{4\pi}}\Delta_0^+\left(- i\partial_{\mu}
+ c_\mu\right) \Delta^a
\epsilon^{\mu \alpha \beta}
F^a_{\alpha \beta} + c.c. \label{GL}
\end{equation}

 Now we need  no extra constraints.  It is
noteworthy that while the ordinary Chern-Simons terms are completely
general covariant,  potential term (\ref{potential}) 
is invariant only under volume
preserving diffeomorphisms.  This should not be disturbing, however,
because there is a preferred density -- though not a preferred shape
-- associated with incompressible quantum Hall liquid. The
non-asymptotic form, with $\lambda_{1,2}$ finite, allows in principle for
the description of localized vortex configurations.

Because the flux-attachment only affects the `trivial' $U(1)$ part of
the topological properties of the Pfaffian state, the
quasiparticle statistics, degeneracy on
the torus, {\it etc.} can be inferred
from those at $\nu=1$.

It is straightforward to generalize this construction to
include more general filling factors. Essentially, all that is required is to
attach an even number of flux quanta in addition to the procedure we used to
map bosons to fermions. This is the standard procedure that is
followed to generate the Jain fractions in the fermionic version of the
abelian Chern-Simons theory of the FQHE \cite{lf}. There is,
however, a significant subtlety concerning the global consistency of
the implementation of 
flux attachment commonly used in the condensed matter literature, 
when it is applied to closed surfaces
of non-trivial topology.  Indeed, the action for the Abelian
gauge field is ordinarily written with
a Chern-Simons term with a coefficient $\theta=1/(2\pi m)$,
where $m$ is an even integer. This is inconsistent with the requirement of
quantization of the Chern-Simons coupling constant needed for a Chern-Simons
gauge theory on a closed manifold \cite{hosotani}. We
should instead take this coefficient to be $1$ and put
the coefficient $1/\theta=(2\pi m)$ in front of the term
(\ref{operator}) which couples the two gauge fields.
In the Appendix we give a more careful
discussion of this point, 
and derive the proper procedure from first principles.

\section{Discussion}

1. A main motivation for our effective field theory is
to establish the existence of a universality class of quantum
Hall states of which the Pfaffian state (\ref{qpfaffian})
is a representative. The most salient property of this
universality class is a $2^{n-1}$-fold degenerate set of
$2n$-quasihole states which  transform as the
spinor representation of $SO(2n)$ as the quasiholes
wind about each other. This is a bit worrisome since
arbitrary perturbations might be expected to
break this degeneracy. In particular, impurities or
small variations in the inter-electron interactions
could, potentially, do this, thereby spoiling the
non-Abelian statistics. With the effective field theories
(\ref{bosoniceft}) or (\ref{fermioniceft}) in hand,
however, it is clear that this degeneracy is, indeed,
stable in the long-wavelength limit. The leading perturbations
are Maxwell terms of the form ${F_{\mu\nu}^a}{F^{\mu\nu a}}$
which are irrelevant by one power of $q$ or $\omega$
compared to the Chern-Simons terms. Perturbations
which couple to the charge density, such as a random potential
or Coulomb interactions, couple to the field strength of the
$U(1)$ gauge field in (\ref{fermioniceft}). Hence,
(\ref{fermioniceft}) is just as stable as an Abelian quantum
Hall state, at least as far as such perturbations
are concerned.

2. A Landau-Ginzburg theory of the Pfaffian state should take
a paired order parameter as its starting point. Indeed, this
is strongly suggested by the appearance of the fields  $\Delta_0, \Delta^a$
in the gauge-invariant form of the effective
field theory (\ref{GL}). This
leads one to the Landau-Ginzburg theory of a superconductor,
but with an Abelian Chern-Simons field to cancel the magnetic
field or, in other words, to a theory almost identical
to the theory of a Laughlin state (\ref{langinz}).
The crucial difference is that
the paired order parameter must have a structure which
allows the existence of neutral fermionic modes at vortex cores, 
where the gap to an unpaired fermion
vanishes. The reader may recall that Bogoliubov-de Gennes
quasiparticles in a superconductor become neutral fermions on  
the Fermi surface and are therefore natural candidates for
the zero modes we are looking for.
We could then, as in \cite{mr,rr,haldane},
interpret the $2^{n-1}$-fold degeneracy
in terms of the occupation of fermionic zero modes
associated with the vortices. This raises the
interesting prospect
that the duality transformation between the
Landau-Ginzburg theory -- in which the electrons
are the fundamental objects -- and the dual theory
of equation (\ref{bosoniceft}) or (\ref{fermioniceft})
-- in which the quasiparticles are fundamental -- is
a highly non-trivial transformation relating
two theories which, at first glance, appear to
be radically different.

3. The observed plateaus at $\nu=5/2$ (in other words,
$\nu=1/2$ in the second Landau level) in a single-layer
system and $\nu=1/2$ in a double-layer system are promising
hunting grounds for excitations with non-Abelian statistics.
One way to experimentally determine whether
either one is described by the Pfaffian state (\ref{pfaffian})
is to measure the quasihole statistics. To do this, we would like
to observe how the quantum state of the system
transforms when quasiholes are braided.
An elegant way to do this utilizes the two
point-contact interferometer proposed by
Chamon, {\it et.\ al.\/} \cite{chamon}.

\begin{figure}
%\centerline{\psfig{figure=interfere2.eps,height=2in}}
%\vskip 0.5cm
\vspace{.2cm}
\noindent
\hspace{1.5 in}
\epsfxsize=4in
\epsfbox{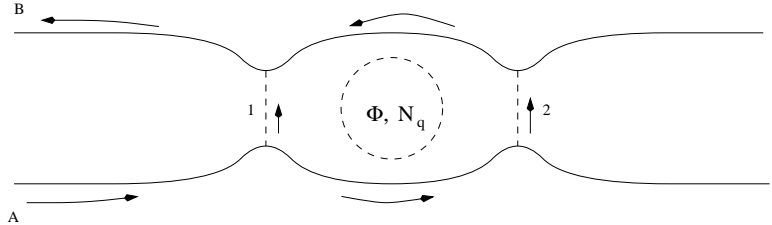}
\vspace{.5cm}
\caption{In the two point-contact interferometer of
Chamon, {\it et.\ al.\/}, quasiholes can
tunnel from the lower edge to the upper edge
by one of two interfering paths. The interference
is controlled by varying the flux, $\Phi$, and number
of quasiholes, $N_q$, in the central region.}
\label{interfere}
\end{figure}

\begin{figure}
%\centerline{\psfig{figure=interwilson3.eps,height=3in}}
%\vskip 0.5cm
\vspace{.2cm}
\noindent
\hspace{1.5 in}
\epsfxsize=3.5in
\epsfbox{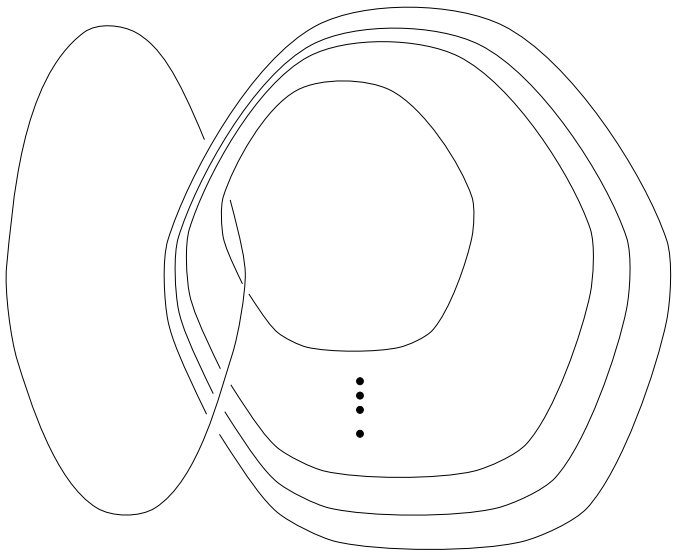}
\vspace{.5cm}
\caption{The matrix element $\langle\psi| {B_{N_q}}|\psi\rangle$
is the expectation value of these Wilson lines.
The loop on the left represents a quasihole which
tunnels at the second point contact, thereby
encircling the $N_q$ quasiholes on the right.}
\label{interwilson}
\end{figure}

In this device (see figure \ref{interfere}),
quasiholes injected at point A along the bottom edge of
the quantum Hall bar can tunnel to
the other edge at either of two point-contacts.
A quasihole which tunnels at the second
point-contact will follow a path which encircles
a central region containing flux $\Phi$ as well as
$N_q$ quasiholes. As a result of the flux
$\Phi$, it will aquire an Aharonov-Bohm phase
which depends on its fractional charge.
Its state will also be tranformed because
its trajectory braids the $N_q$ quasiholes
in the central region. Consequently, the
interference between the two tunneling paths
will depend on the charge and statistics
of the quasiholes. As discussed by Chamon, {\it et.\ al.\/} \cite{chamon},
if we hold the electron number (and therefore the
quasihole number) in the central region fixed,
then the conductance will oscillate as a function
of $\Phi$ with period $\frac{e}{e^*}\Phi_0$,
where $e^*$ is the quasihole charge. If, on the other
hand, we vary $N_q$, we can probe the statistics.
Let us suppose that a quasihole which is injected at
point $A$ on the bottom edge in figure \ref{interfere}
and tunnels at the first point-contact arrives at point B
in state $|\psi\rangle$ and a quasihole which
tunnels at the second point contact is in
the state ${e^{i\alpha}}\,{B_{N_q}}|\psi\rangle$,
where ${B_{N_q}}$ is the braiding operator for
the quasihole to encircle the quasiholes in
the central region and ${e^{i\alpha}}$
is the additional Aharonov-Bohm and dynamical
phase aquired along the second path. Then
the current which is measured at $B$
will be proportional to
\begin{equation}
\frac{1}{2}\left({|{t_1}|^2} + {|{t_2}|^2}\right)\,+\,
{\rm Re}\left\{ {t_1^*}{t_2}\, {e^{i\alpha}}\,
\langle\psi| {B_{N_q}}|\psi\rangle \right\}
\end{equation}
where $t_1$ and $t_2$ are the tunneling amplitudes
at the two point-contacts.
Now, the matrix element
$\langle\psi| {B_{N_q}}|\psi\rangle$
is given precisely by the expectation value
in the effective field theory (\ref{bosoniceft})
of the Wilson lines of figure \ref{interwilson}
or, simply, by the Jones polynomial
${V_{N_q}}({e^{\pi i/4}})$ of these loops.

In other words, if the Pfaffian
state exists in nature, a two point-contact interferometer
will measure the Jones polynomial! By studying the
dependence on $N_q$ -- the conductance will
exhibit the periodicities of the two eigenvalues
of ${B_{N_q}}$ -- we will be able to extract
the non-Abelian statistics.

\section{Acknowledgements}

This work was supported in part by the NSF, grant numbers
NSF DMR94-24511  at UIUC, and NSF PHY94-07194 at ITP-UCSB. EF
was a participant at the ITP Program on {\it Quantum Field Theory in
Low Dimensions}, and thanks the Director of ITP for
his kind hospitality. C. N.  would like to thank  ICTP, Trieste
for hospitality during the workshop on {\it Open Problems
in Strongly Correlated Systems}, where part of this
work was done. A. M. T. is grateful to I. I. Kogan for interesting
discussions. 

\begin{appendix}

\section{}

We will show how to define flux attachment in a manner compatible with the
requirement of quantization of the abelian Chern-Simons coupling constant
or, what is the same, of invariance under large gauge tranformations.
This issue does not arise for the non-abelian sector since its coupling
constant is already correctly quantized.

Consider a theory of particles (in first
quantization) which interact with each other as they evolve in time. We
will assume in what follows that the particles are fermions (in two
spacial dimensions) and that their worldlines never cross.
The actual choice of statistics is not important in what follows but
the requirement of no crossing is important and, for bosons, it implies
the assumption that there is a hard-core interaction while for fermions
the Pauli principle takes care of this issue automatically. For simplicity,
we will assume that the time evolution is periodic, with a very long period.

The worldlines of the particles can be represented by a
conserved current $j_\mu$. For a given history of the system, the
worldlines form a braid with a well defined linking number $\nu_L[j_\mu]$,
given by
\begin{equation}
\nu_L[j_\mu]= \int d^3x \; j_\mu(x) B^\mu(x)
\label{jdotB}
\end{equation}
where $j_\mu$ and $B_\mu$ are related through Amp{\`e}re's Law
\begin{equation}
\epsilon_{\mu \nu \lambda} \partial^\nu B^\lambda(x)=j_\mu(x)
\label{ampere}
\end{equation}
Under the assumption of the absence of crossing of the worldlines of the
particles, the linking $\nu_L[j_\mu]$ is a topological invariant.

Thus, if $S[j_\mu]$ is the action for a given history,
the quantum mechanical amplitudes of all physical observables
remain unchanged if the action is modified by
\begin{equation}
S[j_\mu] \to S[j_\mu]+2\pi  n \nu_L[j_\mu]
\label{shift}
\end{equation}
where $n$ is an arbitrary integer.

The quantum mechanical amplitudes are sums over histories of the
particles, and take the form
\begin{equation}
{\rm Amplitude} \propto
\sum_{[j_\mu]} e^{i S[j_\mu]+2\pi i n \nu_L[j_\mu]} e^{i\phi[j_\mu]}
\label{sum}
\end{equation}
where $\phi[j_\mu]$ is a phase factor which accounts for the statistics of
the particles.

However, the amplitudes remain unchanged if the integrand
of Eq.~\ref{sum} is mulitiplied by $1$ written as the expression
\begin{equation}
1 \equiv \int {\cal D} b_\mu \; \prod_x
\delta(\epsilon_{\mu \nu \lambda} \partial^\nu b^\lambda- j_\mu)
=
{\cal N} \int {\cal D} b_\mu \; {\cal D} a_\mu \;
\exp \left( {\frac{i}{2\pi}} \int d^3x \; a^\mu
\left[\epsilon_{\mu \nu \lambda} \partial^\nu b^\lambda-
j_\mu \right] \right)
\label{one}
\end{equation}
where ${\cal N}$ is a normalization constant and we have used a
representation of the delta function in terms of a Lagrange multiplier
vector field $a_\mu$. Notice that, since $j_\mu$ is locally conserved
({\it i.\ e.\/} $\partial_\mu j^\mu=0$), these expressions are invariant under
the gauge transformations $a_\mu(x) \to a_\mu(x) +\partial_\mu
\Lambda(x)$.

After using the constraint $j=\partial \wedge b$, the amplitude can also be
written
in the equivalent form
\begin{equation}
{\rm Amplitude} \propto
\sum_{[j_\mu]} \int {\cal D} b_\mu \; {\cal D} a_\mu \;
 e^{i S[j_\mu]+2\pi i n \nu_L[j_\mu]} e^{i\phi[j_\mu]}
e^{i \int d^3x a^\mu (x) {\frac{1}{2\pi}}
\left[\epsilon_{\mu \nu \lambda} \partial^\nu b^\lambda-
j_\mu \right]}
\label{amp1}
\end{equation}
We can then compute this amplitude as a path integral of a theory in
which the particles whose worldlines are represnted by the currents
$j_\mu$, interact with the gauge fields $a_\mu$ and $b_\mu$. These
interactions are encoded in the effective action
\begin{equation}
S_{\rm eff}[a,b,j]= {\frac{1}{2\pi}} a^\mu
\left(\epsilon_{\mu \nu \lambda} \partial^\nu b^\lambda-
j_\mu\right) + {\frac{2n}{4\pi}} \epsilon_{\mu \nu \lambda}
b^\mu \partial^\nu b^\lambda
\label{effective}
\end{equation}
where we have solved the constraint $j=\partial \wedge b$ to write the
term of the winding number in the form of a Chern-Simons action for the
gauge field $b_\mu$ ~\cite{wz,wuzee}. Hence, the amplitudes can be
written in terms of a path integral over an abelian Chern-Simons gauge
field with a correctly quantized coupling constant equal to
${\frac{2n}{4\pi}}$.

The usual form of the flux-attachment transformation is found by
integrating out the gauge field $b$. For vanishing boundary conditions
at infinity, this leads to an effective action for the field $a_\mu$ of
the conventional form~\cite{lf}
\begin{equation}
S_{\rm eff}[a]={\frac{1}{2}} {\frac{1}{2\pi 2n}} \int d^3x
\epsilon_{\mu \nu \lambda} a^\mu \partial^\nu a^\lambda
\label{Sa}
\end{equation}
This form of the effective action is not valid for manifolds with
non-trivial topology. However, Eq.~\ref{effective} is correct in all cases as
it
is both invariant under both local and large gauge transformations.
Notice that the gauge field $b_\mu$ is the dual field referred to in the
rest of the paper and it plays a central role in Wen's construction of
the Abelian FQH hierarchy~\cite{wen}.

\end{appendix}

%\begin{figure}
%\centerline{\psfig{figure=skein3.eps,height=1.5in}}
%\vskip 0.5cm
%\caption{The loops (a) $\gamma$ (b) $\gamma'$ and (c) $\gamma''$
%which enter the skein relation. The three loops differ only
%by the braiding shown. The common shaded section
%is arbitrary.}
%\end{figure}

%\begin{figure}
%\centerline{\psfig{figure=interfere2.eps,height=2in}}
%\vskip 0.5cm
%\caption{In the two point-contact interferometer of
%Chamon, {\it et al.}, quasiholes can
%tunnel from the lower edge to the upper edge
%by one of two interfering paths. The interference
%is controlled by varying the flux, $\Phi$, and number
%of quasiholes, $N_q$, in the central region.}
%\label{interfere}
%\end{figure}

%\begin{figure}
%\centerline{\psfig{figure=interwilson2.eps,height=3in}}
%\vskip 0.5cm
%\caption{The matrix element $\langle\psi| {B_{N_q}}|\psi\rangle$
%is the expectation value of these Wilson lines.
%The loop on the left represents a quasihole which
%tunnels at the second point contact, thereby
%encircling the $N_q$ quasiholes on the right.}
%\label{interwilson}
%\end{figure}

\newpage

\end{document}